\documentclass[letterpaper,preprintnumbers,floats,twocolumn,prd,aps]{revtex4}
\usepackage{amssymb,amsmath}
\usepackage{graphicx}

\begin{document}

\title{
$~$\\ Gravitational Instability of de Sitter Compactifications
}
\author{Carlo~R.~Contaldi$^{(1)}$, Lev Kofman$^{(1)}$, and Marco
Peloso$^{(2)}$}
\address{$^{(1)}${\it CITA, University of Toronto, Toronto, ON,
Canada, M5S 3H8,}}
\address{$^{(2)}${\it School of Physics and Astronomy, University of
Minnesota, Minneapolis, MN 55455, USA.}}

\date{\today}

\begin{abstract} 

We consider warped compactifications in $(4+d)$-dimensional theories,
with four dimensional de Sitter $dS_4$ vacua (with Hubble
parameter H) and with a compact internal space. After introducing a
gauge-invariant formalism for the generic metric perturbations of
these backgrounds, we focus on modes which are scalar with respect to
$dS_4$.  The physical eigenmasses of these modes acquire a large
universal tachyonic contribution $-12d/(d+2) H^2$, independently of
the stabilization mechanism for the compact space, in addition to the
usual KK masses, which instead encode the effects of the
stabilization. General arguments, as well as specific examples, lead us to
conjecture that, for sufficiently large dS
curvature, the compactified geometry becomes gravitationally unstable
due to the tachyonic growth of the scalar perturbations.  This mean that for
any stabilization mechanism the
curvature of the dS geometry cannot exceed some critical value.  We
relate this effect to the anisotropy of the bulk geometry and suggest
the end points of the instability. Of relevance for inflationary
cosmology, the perturbations of the bulk metric inevitably induce a
new modulus field, which describes the conformal fluctuations of the 4
dimensional metric.  If this mode is light during inflation, the
induced conformal fluctuations will be amplified with a scale free
spectrum and with an amplitude which is disentangled from the standard
result of slow-roll inflation. The conformal 4d metric fluctuations
give rise to a very generic realization of the mechanism of modulated
cosmological fluctuations, related to spatial variation of couplings
during (p)reheating after inflation.

\end{abstract}

\preprint{CITA-2004-04}
\maketitle


\section{Introduction}

De Sitter or quasi-de Sitter 4d geometries describe the present day
acceleration of the Universe as well as the inflationary expansion at
very early times.  This stimulates significant interest towards
the construction of 4 dimensional dS geometry in the context of
fundamental string/M theory, which are formulated in 10/11 dimension,
with a compact internal space \cite{Mal,ds1,KKLT}.  Similarly, compactification
to 4d dS geometry takes place in phenomenological braneworld models,
where the inner space is periodic with orbifold branes at the edges
\cite{ds2}. Although most of the activity in this area has been
enhanced by very recent progress (both on the observational and
theoretical side) the issue of dimensional reduction to the outer
cosmological space--time was popular since the 1980s, either in high
dimensional supergravity theories or on phenomenological grounds. For
example, see the collection of references on KK cosmology given in
\cite{ss}.

The geometries we discuss here have $d$ spatial dimensions wrapped on
a compact manifold ${\cal M}\,$, in addition to the standard $(3+1)$
space--time.  Many of the mentioned examples are covered by the $(4+d)$
dimensional geometry with the metric
\begin{eqnarray}\label{blocks}
d s^2 = e^{2\hat A(y)}\left(-dt^2+e^{2Ht} d \vec x^2\right)+e^{-2\hat A(y)}{\hat g}_{a b} \left( y \right) d y^ad y^b \,,
\end{eqnarray}
where the outer space is a dS 4d metric with a Hubble parameter $H$,
while ${\hat g}_{a b}$ is the metric of the compact inner space with
$d$ coordinates $y^a$. For generality, we also include the warp factor
$\hat A(y)$. In the following we use greek letters to describe the
$(3+1)$ outer space--time coordinates ($\mu=0,...,3$) while roman
letters span the inner compact space coordinates only. 
Capitalized roman letters span all coordinates.
With the re-definition of the inner space metric ${ g}_{a b} \equiv
e^{-4\hat A(y)}{\hat g}_{a b}$ and the warp factor $A(y) \equiv
e^{\hat A(y)}$, we can re-write (\ref{blocks}) in a
conformally-factorized form
\begin{equation}\label{conf}
d s^2=A(y)^2 \left(d s_4^2 +{ g}_{a b} d y^a \, d y^b \right) \ ,
\end{equation}
where $ds_4^2$ is de Sitter metric. This form of the metric is more
convenient for developing the formalism of metric perturbations around
a warped $dS_4 \times {\cal M}\,$ background, which we will present
below.

>From the string theory/supergravity perspective, recent studies have
concentrated on the compatibility of high $(4+d)$ dimensional geometries
($d=6$) with a 4d de Sitter geometry as the outer space--time, and on
the stabilization of the internal space to $ dS_4$ \cite{KKLT}. The
bulk geometry, which is usually treated in the supergravity limit,
requires a careful study of the $(4+d)$ dimensional (bulk) Einstein
equations.  The progress in finding solutions is related to
the identification of various possible sources for the bulk stress energy
tensor $T^{A}_{B}$, including supergravity lagrangian fields, branes
with fluxes, etc. In phenomenological braneworld models, the $(4+1)$ or
$(4+2)$ dimensional bulk geometry is stabilized, for instance, by means
of bulk scalar fields as in the Goldberger-Wise model~\cite{GW}, or
more generally through the Casimir effect \cite{Cas,pp}.

Very often, however, the stabilization is studied
at the level of a 4 dimensional effective theory, where the inner
space geometry emerges in terms of moduli fields. If the effective
potential of the moduli has a minimum, the moduli are considered to be
stabilized. 
An example will be the stabilization mechanism with quantum field
theory effects which emerge from the properties of ${\cal M}\,$, e.g.
the Casimir effect from compact inner dimensions. 
It is a reasonable assumption that in principle there shall be a bulk
prototype  $\langle T^{A}_{B}\rangle$
(may be as a complicated functional of the
metric and the topology) which generates corresponding terms of the 4d
effective potential. 
The version of  string theory dimensional reduction of~\cite{KKLT} 
is realized with the
participation of the instanton effects (which are manifested in 4dim
superpotential of $N=1$ supergravity).
An interesting question is
whether it is possible to think about a bulk
prototype  $T^{A}_{B}$  of these effects.

While it is reasonable to assume that the low energy 4d
effective action is Einstein gravity plus moduli fields (or Brans-Dike
gravity plus moduli), this picture lacks the full higher dimensional
solution of the Einstein equations with proper sources. In particular,
the four dimensional description is inadequate for studying the high
energy regime (high dS curvature) where the stabilization can  break down.
For example, in the 5 dimensional braneworld models, the issue of
stabilization due to the bulk scalar field can be formulated fully in
terms of the stability of the 5 dimensional warped geometry against
scalar metric perturbations. From this study one can determine the
modes which enter in the exact low energy 4 dimensional description,
namely the radion and the other KK modes of the system \cite{radion},
which cannot be found from an heuristic 4 dimensional effective
potential.

We are not concerned here with the details of the stabilization
mechanism.  Our main goal is to determine the effects of the dS
expansion on the stability of these geometries.  As we will show,
compactifications to a 4dim dS space--time are more difficult to
achieve than compactification to a 4dim flat space--time. This is due to
a tachyonic contribution to the square of the mass of scalar metric
perturbations arising from the de Sitter curvature.  In what follows
we will discuss the properties of $(4+d)$ dimensional classical Einstein
equations assuming a bulk $T^{A}_{B}$ as a source, but without
specifying it.  The gravitational instability which we will discuss
comes from the gravity sector, so that the exact form of $T^{A}_{B}$
will not be crucial. Of course it will be attractive to check how the
effect works for each particular model of $T^{A}_{B}$.  However, we
will try to argue that the instability effect is generic.

Previous studies of the gravitational sector ({\it i.e.} metric
perturbations) of the compactification concentrated on the dimensional
reduction to 4dim flat space--time \cite{ss}.  The instability effect
which we find emerges when the outer space is curved instead of flat, and
it is proportional to the curvature. Notice that test scalar fields propagating in
$dS_4 \times {\cal M}$ do not exhibit tachyonic masses under KK projection~\cite{KS}.
Thus, there is a significant difference in the mass spectrum between the
test fields and the self-consistent treatment of the metric
fluctuations around this background.

The plan of this paper is as follows. In Section ~\ref{sec:metric} we
introduce a generic formalism of metric fluctuations around the $dS_4
\times {\cal M}$ background and define scalar, vector and tensor
fluctuations acording to their transformation properties relative to
$dS_4$. In Section~\ref{sec:lin} we introduce the linearized equations
that govern the evolution of scalar perturbations of the metric and
show how the tachyonic instability of the scalar modes arises from the
gravitational sector. In Sections~\ref{sec:ex1} and ~\ref{sec:ex2} we
show two explicit example where the instability is manifested. In
Section~\ref{sec:mod} we discuss how the instability may be relevant for
the generation of the primordial perturbations, providing a very natural and
model-independent realization of the idea of modulated
perturbations~\cite{modulated,dgz}. Conclusions and a brief discussion can be
found in Section~\ref{sec:end}.

\section{Metric perturbations around the bulk geometry}\label{sec:metric}


The setting of the problem we are discussing is similar to the routine setup
for $(3+1)$ dimensional cosmology with the metric $ds^2=a(t)^2
\left(-dt^2+d \sigma_3^2 \right)$, where $d \sigma_3^2$ is the metric
of 3d space of constant curvature.  Perturbations of this metric can
be classified in terms of freely propagating gravitational waves, together with
vector and scalar perturbations, supported by the perturbations of the
sources (for instance scalar field fluctuations $\delta \phi$ in the
scalar field cosmology).  The classification of cosmological
fluctuations is based on the gauge-invariant Bardeen's variables
\cite{B} (see also \cite{mfb}).

In the background geometry (\ref{conf}) there are also different modes
of metric perturbations, corresponding to different irreducible
representations of the background symmetries.  Modes living in
different representations are not coupled at the linear level. For
this reason, it is very useful to extend the classification
of~\cite{B,mfb} to this case too.

\subsection{General formalism of $(4+d)$ dimensional  metric fluctuations}\label{sec:metric1}

Bearing in mind the similarities in the structures of 
our bulk metric (\ref{conf}) and the $(3+1)$  cosmological metric, 
the generalized perturbed form of the metric (\ref{conf}) can
be written as

\begin{widetext}
\begin{equation}\label{form}
ds^2 = A^2(y)\left\{ { g}_{ac}(\delta^c_b+2\phi^c_b)dy^ady^b
+\left[ (1+2\psi)g_{\mu\nu} + 2E_{;\mu\nu} +
  2F_{(\mu;\nu)}+h_{\mu\nu}\right]dx^\mu dx^\nu - 2A_{a\mu}dy^adx^\mu\right\}
\end{equation}
\end{widetext}

with $g_{\mu\nu}dx^\mu dx^\nu = ds^2_4$ and where $F_\mu$ and
$h_{\mu\nu}$ are divergence free and $h_{\mu\nu}$ is traceless. The
cross component $A_{a\mu}$ can be further decomposed into individual
scalar and vector contributions as $A_{a\mu} = B_{a;i} + S_{ai}$. It
is simple to check that the variables $\phi^c_b, \psi, E, F_{\mu},
A_{a\mu}, h_{\mu\nu}$ account for all possible physical metric
perturbations.

Gauge invariant variables that are counterparts to the Bardeen
variables in $(3+1)$ dimensions can be derived in this setup by
considering infinitesimal coordinate transformations $x^A \rightarrow
{\tilde x}^A = x^A + \xi^A$. In particular we have
new scalar variables
\begin{eqnarray}
  \Phi_{ab} &=& \phi_{ab} + \frac{A^{;c}}{A}(B_c-E_{;c}){
  g}_{ab} - (B_{(a}-E_{(;a})_{;b)}, \\
  \Psi &=& \psi + \frac{A^{;c}}{A}(B_c-E_{;c}).
\end{eqnarray}

In terms of {\it four} dimensional tensor properties in the 4d
observable world, we distinguish scalar perturbations $\phi^c_b, \psi,
E$, vector perturbations $F_{\mu}, A_{a\mu}$ and tensor perturbations
$ h_{\mu\nu}$. 

This is a rather important and subtle point. What is a scalar
fluctuation with respect to 4 dimensional de Sitter space--time, may
have a non--scalar nature in the bulk geometry. For instance, a free
graviton in the bulk may have a gravi--scalar projection in the 4d
space--time. Below we will construct a specific example where the
scalar metric perturbation of (\ref{form}) in fact comes from free
bulk graviton.

Our grouping of the metric perturbations (\ref{form}) is very similar
to the metric fluctuations formalism of \cite{bruc} designed for
higher dimensional braneworlds.  The important distinction between our
formalism and that of \cite{bruc} are the following: we keep the time
coordinate in the homogeneous part of the de Sitter metric, and the
tensorial properties of fluctuations are related to the 4 dimensional
symmetric part. In \cite{bruc} the tensorial properties are related to
the 3 dimensional homogeneous space. In addition; we do not rely in
any way on the existence of branes in the models. On the other hand,
our formalism can
embrace the braneworld inflation as a particular case.

As mentioned, the main advantage of using the split (\ref{form}) is
that perturbations of different types are not coupled at the linear
level. The analysis is further simplified if the same decomposition is
performed also for the perturbations of the source term $\delta
T^A_B\,$. In this work we study metric perturbations of the scalar
type.
 \footnote{Although we do not investigate tensor modes $h_{\mu\nu}$ in
   this work, in passing we can make the following remark: it is
   known that tensor modes on the dS background have the zero KK mode
   followed by a gap of amplitude $2H^2$ due to the unitary constraint
   \cite{Higuchi}. Also, notice that $h_{\mu\nu}$ does not describe all
   the projections of the bulk graviton.}

\subsection{Scalar perturbations}

For this case we can define a generalized longitudinal gauge analogous
to the $(3+1)$ dimensional case. In this gauge the scalars $B_a$ and $E$
vanish, and the scalar perturbations $\phi_{ab}$ and $\psi$ coincide
with the gauge invariant variables $\Phi_{ab}$ and $\Psi$,
respectively.  Thus, without lost of generality, the `scalar' part of
metric fluctuations around a geometry with $dS_4 \times {\cal M}\,$
symmetry can be simply written as
\begin{equation}
ds^2 = A^2(y)\left\{ { g}_{ac}(\delta^c_b+2\phi^c_b)dy^ady^b
+(1+2\psi)ds^2_4\right\}
\end{equation}

We can obtain a further reduction in the degrees of freedom
where the internal space ${\cal M}$ is maximally symmetric. In this case
$\phi^a_b \rightarrow \delta^a_b \phi\,$, and we obtain
\begin{eqnarray}
d s^2 = A^2 \left( y \right) \left[ \right. \!\!\!\! && \!\!\!\! \left(
1 + 2 \, \Psi \left( x, y \right) \right) d s_4^2 + \nonumber\\
&&\left. + \left( 1 + 2 \, \Phi \left( x, y \right) \right) {
g}_{a b} \left( y \right) d y^a \, d y^b \right] \,\,.
\label{line}
\end{eqnarray}
Eq.~(\ref{line}) is the the starting
point for our treatment of scalar perturbations
It is instructive to compare the simple form of scalar metric
fluctuations in the longitudinal gauge around a $(4+d)$ symmetric
$dS_4 \times {\cal M}\,$ background with lower dimensional examples.
If we replace $A(y) \to a(t)$, $ d s_4^2 \to d \sigma_3^2$, $g_{a b} d
y^a \, d y^b \to -dt^2$, eq.~(\ref{line}) reduces to the familiar form
of scalar metric perturbations around a $(3+1)$ dimensional FRW cosmology.
Secondly, if $g_{a b} d y^a \, d y^b=dy^2$ is simply one extra dimensional
line element, eq.~(\ref{line}) gives us scalar metric perturbations in
the $(4+1)$ braneworlds \cite{cgr,radion}, where $\Phi$ is related to
the radion.


\section{Linearized Equations for Scalar Perturbations}\label{sec:lin}

First we perform a customary perturbative analysis of the linearized
Einstein equations for the scalar metric perturbation~(\ref{line}). It is
remarkable that, for symmetric ${\cal M}$, the two scalars $\Phi$ and
$\Psi$ completely describe all the scalar metric fluctuations around
dS $\times$ ${\cal M}\,$. The modes $\Phi$ and $\Psi$ will
be coupled to scalar perturbations of $\delta T^A_B\,$, if they are
present. Typical example of scalar fluctuations in $\delta T^A_B$ are
the fluctuations of the bulk scalar field, which is used in the 
stabilization of ${\cal M}\,$.

We will show that $\Phi$ and $\Psi$ are proportional to each other,
and decompose $\Psi$ into scalar harmonics (KK modes) on $dS_4$. We
will then construct the general expression for the lowest eigenvalue
(square of the physical mass) of the scalar harmonics, and show that it
contains a tachyonic contribution, which signals a possible
gravitational instability.

 It is very important to stress that $\Phi$ and $\Psi$ exist
as dynamical fields even if $\delta T^A_B$ is absent, contrary to the
case of $(3+1)$ dimensional cosmology.  Indeed the bulk geometry dS
$\times$ ${\cal M}\,$ is anisotropic as a whole. Therefore we may
expect the presence of an anisotropic mode of the gravitational
instability, even when the fluctuation of the source
$\delta T^A_B$ are absent.  It
turns out that this pure gravitational mode is also manifested in the
scalar fluctuations $\Phi$ and $\Psi$.  In the next section we put the
metric (\ref{line}) into the context of the generalized anisotropic
Kantowsky-Sacks geometries and show that the longitudinal metric
fluctuations around (\ref{blocks}) correspond to gravitationally
unstable modes of the anisotropic metric (\ref{blocks}).

\subsection{Linearized Einstein Equations}

Consider the linearized Einstein equations for the scalar
perturbations $\Phi$ and $\Psi$.  The equations for $\delta G^\mu_\nu$
contain a diagonal part, proportional to $\delta^\mu_\nu$, plus a
second term which is nontrivial also for $\mu \neq \nu\,$. It amounts
to
\begin{equation}\label{off}
\delta G^\mu_\nu =
 - A^{-2} \, \nabla^\mu \, \nabla_\nu \left[ 2 \, \Psi + d \, \Phi \right]  \;\;,\;\;\; \mu \neq \nu  \,\,,
\end{equation}
where $\nabla_\mu$ is the covariant derivative with respect to the
non-compact de Sitter space. In the absence of an anisotropic stress
tensor in the outer space, $\delta T^{\mu}_{\nu}=0$ for $\mu \not= \nu
\,$, (which is typical in cosmology), eq.~(\ref{off}) imposes
\begin{equation}\label{conn}
\Phi = -  \frac{2}{d}  \, \Psi  \ .
\end{equation} 
This formula replaces the well known relation for $\Phi$ and $\Psi$
 scalars in $(3+1)$ dimensional cosmology, $\Phi=-\Psi$ \cite{mfb}, and
generalizes the result $\Phi=- 2 \, \Psi$ valid 
 for the $(4+1)$ braneworlds~\cite{FK}.

The linearized Einstein equations for the scalar perturbations $\Psi$
yield
%
%
\begin{widetext}
\begin{eqnarray}
 \delta G^\mu_\nu &=& A^{-2} \delta^\mu_\nu \left[ 6
\, H^2 + \frac{d+2}{d} \left( { \Box_d} - \frac{2}{d+2} \, { R} + L_1
\right) \right] \Psi = {8\pi G} \, \delta T^\mu_\nu \propto \delta^\mu_\nu   \,\,, \label{linpert1} \\
\delta G^\mu_a &=& - A^{-2} \frac{d+2}{d} \,
\left[ \partial^\mu \partial_a + L_2 \right] \Psi  = {8\pi G} \, \delta T^\mu_a \,\,, \label{linpert2}  \\
\delta G^a_b &=&  A^{-2} \frac{d+2}{d} \left[ -
2 { \nabla}^a { \nabla}_b + \frac{4}{d+2} { G}^a_b +
\delta^a_b \left( \Box_4 + \frac{12 d}{d+2} H^2   + 2 \, { \Box_d}
\right) + L_3 \right] \Psi =   {8\pi G} \, \delta T^a_b\,\,.
\label{linpert}
\end{eqnarray}
\end{widetext}
In these equations $\Box_4 = - \partial_t^2 - 3 H \partial_t + {\rm
e}^{-2 H t} \partial_i^2$ is the scalar $\nabla^\mu \, \nabla_\mu$
operator of the $(3+1)$
dimensional de-Sitter space, while ${\nabla_a} ,\, { G^a_b}= {
R^a_b}-\frac{1}{2}\delta^a_b{ R},\, { \Box_d}$ are computed on the
compact manifold ${\cal M}$ with the metric ${ g}_{a b}\,$.  $L_1,
L_2, L_3$ denote terms proportional to derivatives of $A\,$. Such
terms do not contain second derivatives with respect to $x^\mu$ and
can be neglected in the present discussion. For completeness, they are
reported in Appendix. Finally, $G$ denotes the higher
dimensional Newton constant.

\subsection{Tachyonic Contribution to the Eigenmasses }

  The scalar
perturbations ~(\ref{line}) as well as the scalar perturbations of the
stress-energy tensor $\delta T^A_B\,$, due to the symmetries of the
background geometry can be decomposed into the scalar eigenmodes
$Q_n(x)$
\begin{eqnarray}
\Psi \left( x, \, y \right) &=& \sum_n {\tilde \Psi}_n \left( y \right)
\, Q_n \left( x \right) \,\,, \nonumber\\
\delta  T^A_B\left( x, \, y \right) &=& \sum_n \widetilde{\delta T^A_B}_n
 \left( y \right) \, Q_n \left( x \right) \,\,,
\label{separo}
\end{eqnarray}
where the sum corresponds to the the KK tower of the $4$ dimensional
 scalar modes.  These modes $Q_n$, which are common for all
 fluctuations, obey the free massive scalar field equations
\begin{equation}
\left( - \Box_4 + m_n^2 \right) \, Q_n = 0 \,\,,
\label{eqq}
\end{equation}
where $m_n^2$ are the separation eigenvalues of the decomposition
(\ref{separo}).  Substituting~(\ref{separo}) and~(\ref{eqq})
in~(\ref{linpert}), we obtain a set of equations for each
eigenmode. In practice, we simply replace $\left\{ \Psi ,\, \delta
T^A_B \right\}\,$ by $\left\{{\tilde \Psi}_n ,\, \widetilde{\delta
T^A_B}_n \right\}\,$, and $\Box_4$ by the corresponding eigenvalue
$m_n^2\,$.

Among (\ref{linpert1})-(\ref{linpert}), the equation
(\ref{linpert}) is the dynamical equation while equations
(\ref{linpert1}), (\ref{linpert2}) are the constraint ons.
Moreover, these equations are connected by the Bianchi identity.
Thus, it will be enough to work with the two equations
(\ref{linpert2}) and (\ref{linpert}) only.

Let us consider the dynamical equation~(\ref{linpert}).
We  make the following observations which is the crucial one for
our discussion.
We note that the four dimensional
operator $\Box_4$ only enters in the diagonal components of the
Einstein tensor $\delta G^a_b$
 corresponding to the internal space,
see eq.~(\ref{linpert}).
 Most importantly,  it
always appears in the combination $\Box_4 +
\frac{12 d}{d+2} \,  H^2.$
Hence, this combination will always generate the contribution
\begin{eqnarray}
&& \left( \Box_4 + \frac{12 \, H^2 \, d}{d+2} \right) \Psi
\;\rightarrow\; \mu_n^2 \, {\tilde \Psi}_n^2 \,\,, \nonumber\\
&& \mu_n^2 \equiv m_n^2 + \frac{12 \, H^2}{1+2/d} \,\,
\label{mass}
\end{eqnarray}
in the equations for the modes along ${\cal M}\,$. This is true
for any choice of ${\cal M}\,$ and for any underlying
stabilization mechanism, since it is due to the symmetry of the
background configuration~(\ref{blocks}).
Thus, the eigenvalue $m_n^2$ from the outer space--time 
always appears together with $ \frac{12 \, H^2}{1+2/d}$.

After the factorization (\ref{separo}), the resulting equations 
are  the equations for the functions  $\left\{{\tilde
\Psi}_n(y) ,\, \widetilde{\delta T^A_B}_n (y) \right\}\,$
 on ${\cal M}$. Typically, they can be reduced to a second order elliptic equation.
For compact ${\cal M}$, this  corresponds to the eigenvalue problem
which  determines the
combinations $\mu_n^2\,$.
 Once these values have been
obtained, the physical masses $m_n^2$ of the perturbations immediately
follow from eq.~(\ref{mass}).

Further we can use the symmetry of ${\cal M}$.  If the warping is
absent, $A=const$, on general ground we expect that the solution of
the eigenvalue problem on ${\cal M}$ will depend on the curvature of
${\cal M}$, $\mu_n^2\, \sim R$; we denote it as $\mu_n(R)^2$.  On the
other hand, for non-vanishing warping operators $L$ may bring in
$\mu_n^2\,$ extra dependence from $A(y)$ in terms of its
characteristic scale, which we denote as $\mu_n(A)^2$.

Taking all this together, the eigenvalues of the scalar metric 
perturbations in dS space--time are expected to have the following structure
\begin{equation}\label{mass1}
m_n^2=- \frac{12 \, H^2}{1+2/d} + \mu^2_n(R) +\mu^2_n(A) \ .
\end{equation}
To support this conjecture, in the next Sections below
we will  give examples of  explicit computations of $m_n^2$.

The most important conclusion is that the curvature of non--compact de
Sitter dimensions gives a negative (tachyonic) contribution in
eq.~(\ref{mass1}) to the eigenmasses $m_n^2$ of the scalar
fluctuations.  If de Sitter curvature exceeds a critical value so that
$m_n^2$ becomes negative, we encounter gravitational instability of
the $dS_4 \times {\cal M}$ geometry.  This result generalizes a similar
finding obtained in~\cite{FK} in the case of $(4+1)$ dim braneworlds,
where $m_n^2 = - 4 \, H^2 + \mu_n^2(A)$.  In \cite{FK} the
interpretation of the tachyonic instability in $m_n^2$ was not given.
Here we argue that this is an effect of gravitational instability of
anisotropic geometries $dS_4 \times {\cal M}$.
The present discussion shows that the instability is a generic issue
that is not confined solely to the braneworld models studied
in~\cite{FK,branecode}. Rather, it is a general property of the
geometries of the form~(\ref{line}), which includes a very wide class
of theories with extra dimensions.  In the next Section we discuss how this instability
emerges in different contexts.

\subsection{Tachyonic Instability of $dS_4 \times {\cal M}$}

We now follow what happens if   $m_n^2$ is  negative.
 For this we have to turn to equation (\ref{eqq}) for
the four--dimensional eigenfunction $Q_n(t, \vec x)$.  The
four--dimensional massive scalar harmonics $Q_n$ can be further
decomposed as $Q_n(t,\vec x) = \int f_k^{(n)}(t)\, e^{i \vec k \vec
x}\, d^3k$.  The temporal mode functions $f_k^{(n)}(t)$ obey the
equation
\begin{equation}\label{eq:4}
  \ddot{f} + 3H\dot{f} + \left( e^{-2Ht}k^2 + m_n^2 \right) f = 0,
\end{equation}
where dot denotes time derivative, and we have dropped the labels $k$ and
$n$ for brevity.
We choose the solution of (\ref{eq:4})
which corresponds to the 
positive frequency vacuum-like \footnote{This is bearing in mind the
evolution of the quantum fluctuations during inflation}
 initial conditions in the far past $t
\to -\infty $, $f_k(t) \simeq \frac{1}{\sqrt{2k}} e^{ik\eta}$,
$\eta=\int dt\, e^{-Ht}$. For the tachyonic mode ($m_n^2 <0$), the solution
to equation (\ref{eq:4}) with this initial condition is given in terms
of Hankel functions $f_k^{(n)}(\eta)=\frac{\sqrt{\pi}}{2} H
|\eta|^{3/2} {\cal H}^{(2)}_{\nu} (k\eta)$, with the index
$\nu=\sqrt{\frac{9}{4}+\frac{|m_n^2|}{H^2}}$. The late-time asymptote of
this solution diverges exponentially as $t \to \infty$ ($\eta \to 0$)
\begin{equation}\label{asym}
  f_k^{(n)}(t) \propto  e^{\lambda_n Ht} \ ,
\end{equation}
with the numerical factor
\begin{equation}\label{lambda}
\lambda_n=
\left( {\sqrt{\frac{9}{4}+\frac{|m_n^2|}{H^2}} - \frac{3}{2}} \right) \ .
\end{equation}

The leading term is associated to the lowest negative mass square $m^2_n=m^2_0
\sim - few \times H^2\,$, which
defines the instability of metric fluctuations
\begin{equation}\label{inst}
 \Psi(t, \vec x; y) \propto e^{\lambda_0 Ht}\ .
\end{equation}

In the following sections we consider examples where 
$m_n^2$ can be calculated explicitly,
and we will provide an interpretations of the tachyonic instability.

\section{Example: Instability of $(4+1)$ Inflating Braneworlds}\label{sec:ex1}

 The inclusion of additional fields is typically required to achieve
the stabilization of the internal space. However, this complicates
significantly the study of the perturbations of the system, so that
explicit results have been obtained only for the simplest
configurations. The
theory of metric perturbations was developed for $(4+1)$ Braneworld
models with inflation. Although we do not deal specifically with the braneworlds in
this paper these provide a good example for models that are affected
by the results discussed in this work and constitute 
 an interesting set-up where
calculations have been performed in detail.  Here we consider the $(4+1)$
dimensional braneworld with de Sitter branes of codimension one,
with and without the presence of a bulk scalar field. The geometry of
the system, with the inclusion of scalar perturbations, is again of
the form~(\ref{line}), with parallel branes as edges of the internal
coordinate $y\,$. The function $A \left( y \right)$ is known as warp
factor.

\subsection{Inflating Braneworld without stabilization}

First let us consider an inflating braneworld without stabilization
(i.e. without bulk scalar field).  We assume a negative cosmological
constant in the AdS bulk.  Scalar metric perturbations are given by
the form~(\ref{line}), and from (\ref{conn}) we have $\Phi=- 2 \,
\Psi$.  The Einstein equation for the linearized perturbations $\Phi$
has to be supplemented by the junction conditions at the orbifold
branes, for details see \cite{radion,FK,kmp}.  Equation
(\ref{linpert2}) gives us
\begin{equation}\label{no1}
\Psi'+2 \frac{A'}{A}\Psi=0 \ ,
\end{equation}
where prime stands for $\partial_y$. This relation shows that the KK tower is
absent in this case, and that the only
eigenmode ${\tilde \Psi}_0 \sim 1/A^2$ is present.  Equation
(\ref{linpert}) gives $\left( \Box_4 + 4 H^2 \right) \Psi=0$. We then see that
he perturbation
$\Psi$ is described by the free wave equation.  In other
words, for the branes without stabilization, $\Psi$ is nothing but the
gravi-scalar 4 dim projection of the bulk gravitational wave.  Second,
the square of the mass for this mode is tachyonic
\begin{equation}\label{no2}
m^2_0 = - 4 \, H^2 \ ,
\end{equation}
in exact agreement with our eq.~(\ref{mass1}).
In this case $\mu^2_n(A)=0$, $\mu^2_n(R)=0$.
The braneworld between two de Sitter branes is unstable. The
equation (\ref{no2}) for inflating branes without stabilization was 
obtained in \cite{sasaki}. 

\subsection{Inflating Braneworld with Bulk Scalar  Stabilization}

The addition of a bulk scalar field can, in principle, provide the stabilization
of the braneworld configuration. The scalar field $\phi$ can have a potential
$V$ in the bulk and potentials $U_{1,2}$ at the two branes. The internal
coordinate can be stabilized by suitable choices of these potentials.
However, as we will see, the branes can be curved to such a degree that
the stabilization mechanism will fail.

Let us consider a stationary configuration, characterized by a fixed
interbrane distance, and let us study its stability. Perturbations of
the bulk field $\delta \phi$ are coupled with the scalar perturbation
of the metric, as we described above.  We can decompose them as in
eq.~(\ref{separo}),
\begin{equation}\label{separo1}
\delta \phi \left( x, \, y \right) =\sum_n \widetilde{\delta
\phi}_n \left( y \right) \, Q_n \left( x \right) \,\, .
\end{equation}
Next, we can use the well known expression for the bulk stress energy
tensor $\delta T^A_B$ in terms of $\delta \phi$ and the background field
$\phi(y)$ (see \cite{FK,kmp} for details), and substitute them in the
linearized Einstein equations.  The constraint equation (\ref{linpert1})
gives a connection between $\tilde \Psi_n(y)$ and $\widetilde{\delta
\phi}_n(y)$
\begin{equation}\label{yes1}
(A^2  \tilde \Psi_n)'=\frac{A^2}{3}\phi' \,\, \widetilde{\delta \phi}'_n \ ,
\end{equation}

Equation (\ref{linpert}) yields instead
\begin{equation}\label{yes2}
\left(-\frac{2}{3}\phi'^2 +m_n^2+4H^2 \right)\tilde{\Psi_n}=\frac{\phi'^2}{3A}
\left(\frac{A}{\phi'}\,\, \widetilde{\delta \phi}'_n \right) \ .
\end{equation}
Here, combinations of $A$ and its derivatives are expressed in a more
compact manner through $\phi'$, using the background equations.

Because $\Psi$ and $\delta \phi$ are related by the constraint
equation~(\ref{yes1}), in fact there is only one dynamical degree of
freedom, which is linear combination of the pair. The mass of the four dimensional
modes is most easily obtained from an eigenvalue equation for the wave
functions of the modes along $y\,$.

Combining equations (\ref{yes1}) and (\ref{yes2})
one obtains~\cite{FK}
\begin{eqnarray}
&& \left[ \frac{d^2}{d \, y^2} + m_n^2 + 4 H^2 - \frac{z''}{z} +
\frac{2}{3} \phi'^2 \right] {\tilde u}_n = 0 \,\,, \nonumber\\
&& {\tilde u}_n \equiv \left( A^{3/2} / \phi' \right) {\tilde \Psi}_n
\;\;,\;\; z \equiv \left( A \, \phi' \right)^{-1/2} \,\,,
\end{eqnarray}
 The mass spectrum of the system is determined by the eigenvalue
 problem, i.e. by the solutions of this equation which satisfies the
 boundary conditions for the perturbations at the two branes. It can
 be shown that the lowest eigenmass is~\cite{FK}
\begin{equation}\label{mass3}
m_0^2 = - 4 \, H^2 +\mu_0(A) \ ,
\end{equation}
where $\mu_0(A) \approx \frac{2 \, \int d y/A}{3 \, \int d y/ \left(
 A\, \phi'^2 \right)}$.  From the background Einstein equations we
 have $\phi'^2=6\frac{A'^2}{A^2}- 3\frac{A''}{A}-3H^2$, so that
 $\mu_0(A)$ is expressed through the warp factor and its derivatives.

As we have argued in the general case (set $d=1$ in eq.~(\ref{mass})),
the eigenmasses $m_n$ enter in the Einstein equations in the
combination $m_n^2 + 4 \, H^2\,$. Hence, the $4 \, H^2$ term results
in a negative contribution to the physical masses of the four
dimensional modes. The stabilization mechanism has to be
``sufficiently strong'' to counterbalance this effect, and the
configurations for which the right hand side of eq.~(\ref{mass3}) is
negative are subject to tachyonic instability (\ref{inst}).

Since the warp factor and the expansion rate $H$ both enter in
$\mu_0(A)$, this term should be considered as a functional of $H\,$. Hence
 $H$ affects $m_0^2$ through both terms.  However, $\mu_0(A)$ only
 mildly depends on $H$ and by far, the strongest $H$ dependence is
 given by the negative $- 4 \, H^2$ contribution.  Therefore the
 critical value of dS curvature $H=H_c$ which borders stable and
 unstable brane configurations can be estimated from $4H^2=\mu_0(A)$,
 where $\mu_0(A)$ can be calculated explicitly for concrete braneworld
 models (for specific examples of the calculation of $\mu_0(A)$, see
 \cite{radion,kmp}).

The study of the linear perturbations around the stationary
configuration gives information about its stability, but it cannot be
used to determine the whole dynamics of the system when the
configuration is unstable. The dynamical evolution can be studied
numerically with the {\tt branecode} package, a numerical code
\cite{branecode} designed specifically for this task.
It was found numerically (by methods which are
completely different from the linear analysis described above) that indeed
stationary inflating branes configurations with too high $H\,$ are
unstable.  Moreover, the full numerical treatment of the dynamics
reveals the end points of the non-linear evolution.  It turns out that,
depending on the parameters, unstable brane configurations can be
re-structured to stable inflating branes configurations with lower
$H$ ($H < H_c$).If the parameters of the branemodel do not admit
another stationary configurations, the branes are instead colliding. Bulk
geometry of the collapsing branes asymptotically reaches the 5 dim
Kasner solution $ds_5^2=-dt^2+t^{2p_1}dy^2+\sum_{i=1}^{3}t^{2p_i}dx_i^2$. 
Three of the Kasner indexes $p_i$ are the same, as disctated by the brane
isometries \cite{branecode}.

\section{Example: instability of $dS_4 \times S_d$}\label{sec:ex2}

As the next example we consider a $(4+d)$-dimensional space--time with
 the internal dimensions compactified on a $d-$sphere of radius
 $r_0\,$.  This is a toy model and has no application to compact extra
 dimensions. However it serves as a simple geometry with which we can
 check our conjecture from the GR point of view.  We work out in details
 the case in which the only source term is a cosmological constant
 $\Lambda\,$. The system is rather simple and its dynamics can be
 studied exhaustively (for the low dimension case of $dS_2 \times
 S_2$ see ~\cite{ks}): there is one stationary configuration of the
type~(\ref{line}), which, due to the lack of a stabilization mechanism,
turns out to be unstable. Our main point is that the instability can be
computed precisely with the general approach outlined in Section~\ref{sec:lin}.
Below we will apply this generic formalism to the case of $dS_4 \times S_d$ with
a positive bulk $\Lambda\,$.

However, there is a complementary treatment of stability issue of
$dS_4 \times S_d$ based on the investigation of its gravitational
evolution within a more general class of anisotropic geometries, the
Kantowsky-Sacks metrics, of which $dS_4 \times S_d$ is a particular
solution.  We show that the two approaches are equivalent. This
provides insight on the nature of the tachyonic instability
conjectured for $dS_4 \times {\cal M}$, as the general gravitational
instability in the class of anisotropic metrics. The dynamics of the the
Kantowsky-Sacks metrics in $(4+d)$-dimensions can be analyzed by the
qualitative methods of the dynamical systems The phase diagram can
be extended far from the stationary configuration, so we can find the
end points of the tachyonic instability.

\subsection{Instability of  linear scalar perturbations around  $dS_4 \times S_d$}

We consider the simple situation in which the geometry is factorized
($A=1$ in eq.~(\ref{line})) and the internal manifold ${\cal M}$ is a
sphere of dimension $d\,$ and radius $r_0\,$. We assume that the only
source term is a cosmological constant $\Lambda$. This systems admits
a  stationary solution with static ${S_d}$ and de Sitter
external space. This configuration is characterized by
\begin{equation}
R=3d H^2 \, \;\;,\;\; H = \sqrt{\frac{2 \, \Lambda}{3 \left( d + 2
\right)}} \,\ \ .
\label{dsphere}
\end{equation}
The curvature of the inner space is $R = d(d-1)/r_0^2$. The system
allows for expending  $H >0$ and contracting $H < 0$ phases.

Consider linearized scalar perturbations (\ref{line}) around this
background in the absence of any dynamical field in the system $\delta
T_\mu^\nu = 0$.  The first Einstein equation (\ref{linpert1}) is
satisfied identically due to relation (\ref{dsphere}).  The constraint
equation (\ref{linpert2}) gives us ${\tilde \Psi}\ =const$, so that
only the lightest mode is present; while the tower of KK modes is
absent (this is similar to the case of a braneworld with negative bulk
cosmological constant without stabilization; also in this case the KK tower would
be recovered upon introduction of dynamical fields). The
dynamical Einstein equation (\ref{linpert}) gives us
\begin{equation}
\left( \Box_4  +\frac{12 d}{d + 2}
 H^2  -\frac{4(d-1)}{d(d+2)}R  \right)\Psi=0 \ .
\label{instdsp}
\end{equation}
>From here, we immediately have
\begin{equation}\label{mass5}
m_0^2=- \frac{12 \, H^2}{1+2/d} + \mu^2_0(R)  \ .
\end{equation}
where $ \mu^2_0(R)=\frac{2(d-2)}{d(d+2)}R$.  The term $\mu^2_0(R)\,$,
 proportional to the curvature of the $d-$sphere, has comparable
 magnitude and opposite sign with respect to the tachyonic
 term. However, the net mass squared of $\Psi$ is negative for any
 dimension $d$,
\begin{equation}\label{mass6}
m_0^2=-6H \ .
\end{equation}
This signals the tachyonic instability of the
configuration~(\ref{dsphere}).
Interestingly, the ratio $m_0/H$ does not depend on $d$.
 
Applying eq.~(\ref{inst}) to this example, the linear scalar perturbations
are growing with time according to 
\begin{equation}\label{inst1}
 \Psi(t, \vec x; y) \propto e^{\lambda_0 Ht}={\rm exp}
{\left[\left(\frac{\sqrt{33}-3}{2}\right) \, Ht\right]} \ .
\end{equation}
Notice that (\ref{instdsp}) is the free wave equation; this means that 
$\Psi$ corresponds to the gravi-scalar projection of
the bulk gravitational wave.

\subsection{Gravitational instability of anisotropic geometry $dS_4 \times S_7$.}

\begin{figure*}
\includegraphics[width=11cm]{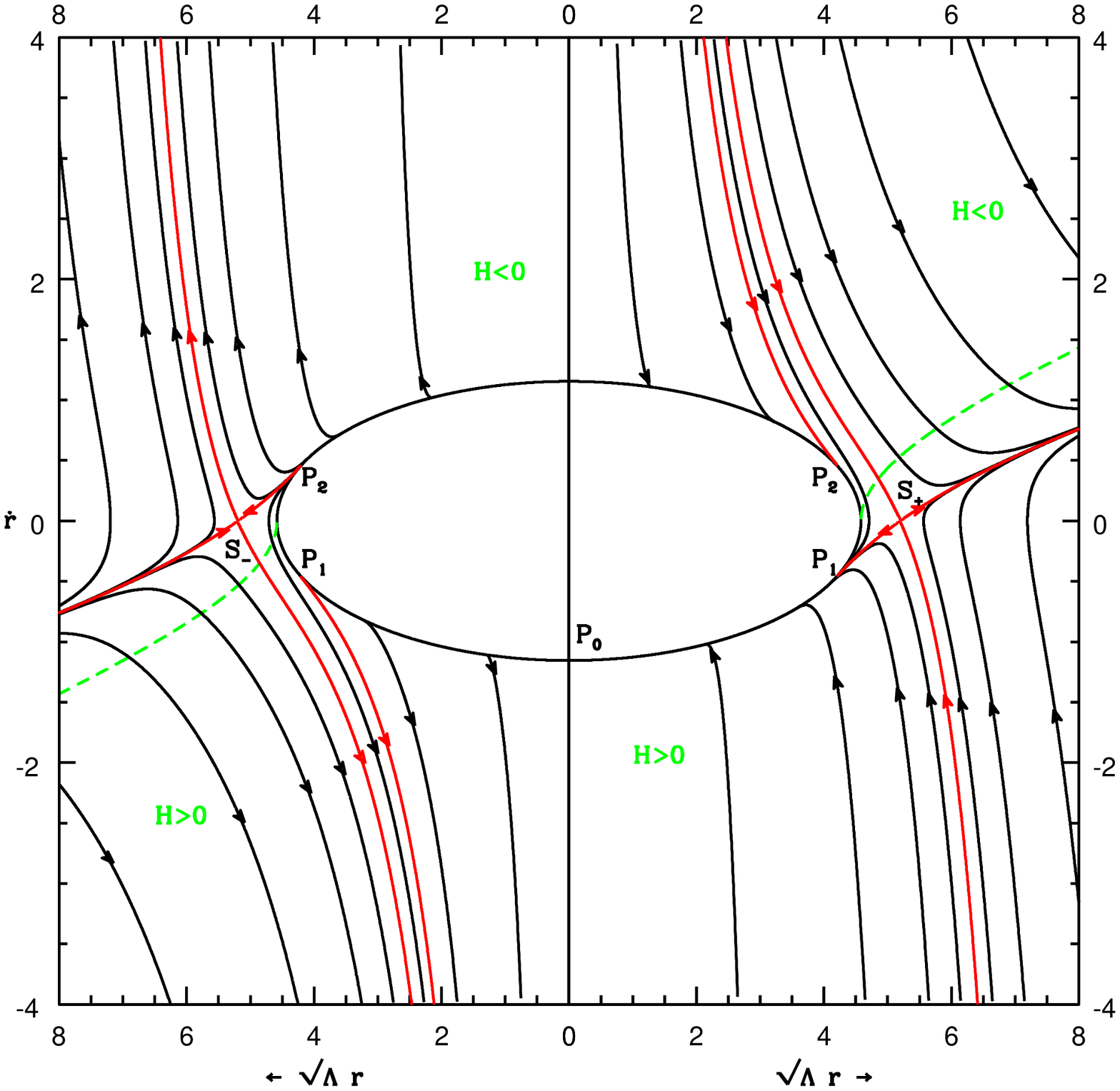} 
\includegraphics[width=5cm]{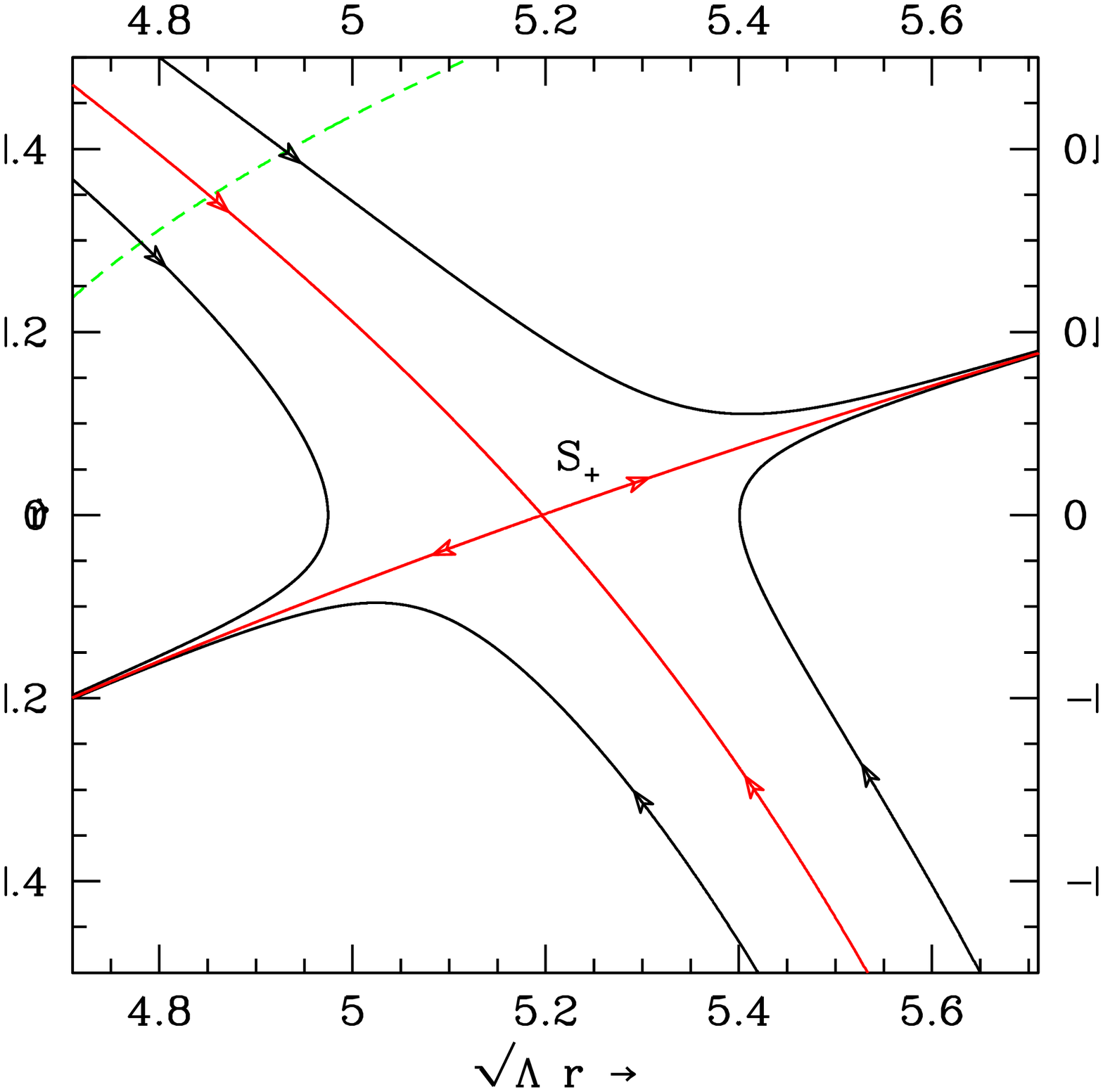}
\caption{\label{phasd}Phase portrait for the dynamical
system~(\ref{dynamical}). Two branches are present, according to
different signs in eqs.~(\ref{dota}), (\ref{dynamical}). The branches
are connected along the boundary of the oval region symmetrically
across the axis passing through the point $P_0$ (as shown by the
identification of points $P_1$ and $P_2$ in each branch. The dashed
(green) curves shown the boundary where the Hubble parameter for the
non--compact space changes sign. The stationary configuration
discussed in the main text is the saddle point $S_+$. Outgoing
trajectories from this point either go to a decompactification of the
internal space (toward asymptotic $dS_{11}$) or they change branch
and go toward Kasner asymptotic with expanding outer space and
collapsing internal manifold. We also show a detailed view of the
critical region around the saddle point $S_+$.}
\end{figure*}

The expression (\ref{inst1})
controls the development of the gravitational instability as long as
$\vert \Psi \vert \ll 1 \,$. However, to understand the nature of the
instability, we have to study the dynamical evolution beyond the
linear regime. To do so, we allow the radius of ${\cal M}$ to be a
time dependent function $r \left( t \right)\,$, and we consider a
general FRW evolution for the non-compact space,
\begin{equation}
d s^2 = - d t^2 + a^2 \left( t \right) d^3 {\bf x} + r^2 \left( t
\right) \left[ d \theta_1^2 + {\rm sin}^2 \theta_1 \, \left( d
\theta_2^2 + \dots \right) \right] \,\,,
\end{equation}
where $\theta_1, \theta_2, \dots$ are angular coordinates on the
$d-$sphere. For definiteness, we specify to $d=7\,$ (the evolution
of the system in a lower
dimensional context was studied in details e.g. in~\cite{ks}). From the
$\left( 0, 0 \right)$ Einstein equation, the expansion rate of the
non-compact dimensions can be written as a function of $r$ and its time
derivative $\dot{r}\,$,
\begin{equation}
\frac{\dot{a}}{a} = \frac{- 21 \dot{r} \pm \sqrt{3} \left[4 \Lambda
    r^2 + 63 \, \dot{r}^2  - 84\right]^{1/2}}{6 \, r} \,\,.
\label{dota}
\end{equation}
We can substitute back this relation into the remaining equations, and
obtain a second order differential equation for $r\,$ alone. As
before, we can rescale $\Lambda$ away by redefining $x \equiv
\sqrt{\Lambda} \, r\,$, and $\tau \equiv \sqrt{\Lambda} \, t \,$. We
then obtain the following dynamical system
\begin{eqnarray}
\label{dynamical}
\frac{d \, x}{d \, \tau} &=& y \,\,, \\
\frac{d \, x}{d \, \tau} &=& \frac{4 x^2 + 81 y^2 -108 \mp 9
\sqrt{3} y \left[4 x^2 + 63 y^2 -84 \right]^{1/2}}{18 \, x} \,\,,
\nonumber
\end{eqnarray}
whose evolution is summarized in the phase portrait of figure~\ref{phasd}.

The set of trajectories lies in a nontrivial manifold, characterized
by a forbidden region ($4 x^2 + 63 y^2 -84 < 0$, bounded area of
fig.~\ref{phasd}), and by two branches which are connected at the
boundary of this region but disconnected anywhere else. The two
branches are shown in the right and left part of fig.~\ref{phasd},
with the right (left) part corresponds to the upper (lower) signs in
eqs.~(\ref{dota}), (\ref{dynamical}). Trajectories are shown by
continuous lines, while the two dashed lines separate the region for
which the external space is expanding ($H>0\,$, lower part in both branches)
from the (upper) one for which it is contracting ($H<0\,$). The trajectories
which reach the boundary of the forbidden region in the right branch
are identified without discontinuity through opposite points along the
boundary into the left branch. The background configuration $dS_4 \times S_7$
studied in the previous Subsection corresponds to the saddle point in the right
branch, characterized by a static radius $\left\{ x=\sqrt{27}, y = 0 \right\}
\, $ (in agreement with eq.~(\ref{dsphere})), and by a de Sitter
outer space. The region of the phase portrait around the saddle point $S_+$
is zoomed in the right panel of fig.~\ref{phasd}.

The linearized system~(\ref{dynamical}) around this
point is
\begin{eqnarray}
\frac{d}{d \, \tau}
\left( \begin{array}{c}
\delta x \\ \delta y
\end{array} \right) =
\left( \begin{array}{cc}
0 & 1\\
\frac{4}{9} & - \sqrt{\frac{2}{3}}
\end{array}
\right) \, \left(
\begin{array}{c}
\delta x \\ \delta y
\end{array}
\right) \,\,.
\label{linearized}
\end{eqnarray}
Next, we have to diagonalize he matrix (\ref{linearized}).
The diagonal components of it are
${\text diag}\left( \frac{-\sqrt{6}-\sqrt{22}}{6},  \frac{-\sqrt{6}+\sqrt{22}}{6}\right)$.
One of the eigenvalue is positive, another is negative, how it shall be for the
saddle point.
Instability corresponds to the positive eigenvalue.
More precisely, metric instability growth with time as ${\rm exp}
\left(\frac{-\sqrt{6}+\sqrt{22}}{6}
\sqrt{\Lambda} t\right)$. Recall that
$\sqrt{\Lambda}=\sqrt{\frac{27}{2}} \, H$ in the case of $S_7$.
Therefore, the metric instability around saddle point grows as
\begin{equation}\label{inst2}
r(t)-r_0 \propto {\rm exp}
{\left((\frac{\sqrt{33}-3}{2}) \, Ht\right)} \ .
\end{equation}
This is precisely the result (\ref{inst1})
 for $\Psi$ obtained with the previous calculation.
This confirms the equivalency of
the two approaches for the study of the instability of the stationary
configuration.

The eigenvalues of the matrix~(\ref{linearized}) are of opposite sign,
indicating that the stationary configuration is a saddle point. Its
eigenvectors describe the critical directions in the linear regime
close to the saddle points. The motion along the unstable line leads
to two very different asymptotics along the two opposite
directions. Starting with a slightly larger $r$ the system goes
toward the decompactification of ${\cal M}$, with the contribution
from the cosmological constant dominating more and more over the
curvature of the $d-$sphere. The asymptotic regime is a straight line
in the phase portrait, $y/x=$ constant. Substituting this ``ansatz''
in eq.~(\ref{dynamical}), one finds $y/x \rightarrow 1/ \sqrt{45}\,$,
as $x,y \rightarrow \infty\,$. This solution corresponds to a de
Sitter expansion of the internal space. From eq.~(\ref{dota}) we see that
the non--compact space also approaches the same de Sitter expansion,
\begin{equation}\label{rate2}
\frac{\dot{a}}{a} \,, \frac{\dot{r}}{r} \rightarrow
\sqrt{\frac{\Lambda}{45}} \,\,.
\end{equation}
Hence, the system tends toward $d S_{11}\,$, with an isotropic
expansion of all the coordinates.
Notice that the bulk expansion rate (\ref{rate2}) is slower than 
the initial expansion rate of the outer space $H=\sqrt{\frac{2\Lambda}{27}}$.
Isotropization of expanding solutions is produces by the 
bulk cosmological constant.

Alternatively, the instability can lead to the collapse of the
internal manifold. In the phase diagram of figure~\ref{phasd}, this possibility
is described by trajectories which, starting from the
stationary point, go from the right to the left branch, and then
continue toward $r \rightarrow 0\,$, while the external space expands
with increasing expansion rate. The asymptotic evolution is of the
Kasner type $ds^2=-dt^2+t^{2p_1}d\vec x^2+t^{2p_4}d\Omega_7^2$. We have 
\begin{equation}
a \propto \left( t_* - t \right)^{-\frac{\sqrt{21}-1}{10}} \;\;,\;\; r
\propto \left( t_* - t \right)^{\frac{3 \sqrt{21} + 7}{70}} \,\,,
\label{kasner}
\end{equation}
where $t_*$ is the time at which the system becomes singular (one can
easily verify that~(\ref{kasner}) solve the dynamical equations at
leading order as $t \rightarrow t_*\,$).
Kasner exponents are the same in two groups,
$p_1=p_2=p_3={-\frac{\sqrt{21}-1}{10}}$ and $p_4=...=p_{11}={\frac{3 \sqrt{21} + 7}{70}}$.
 The Kasner geometry is a
generic collapsing solution~\cite{bkl}, already advocated in string
cosmology~\cite{dh} and braneworld cosmology~\cite{branecode}. It is
characteristic of a strong gravity regime, where the presence of any
source (the cosmological constant, in the case at hand) can be
neglected.

\section{Modulated perturbations}\label{sec:mod}

In the previous Sections we discussed how the de Sitter expansion can
affect the stability of the internal space. As we discussed, the
stability cannot be reached by gravity alone. A point of
phenomenological relevance is that the field(s) responsible for the
stabilization unavoidably introduce (quantum) fluctuations which
perturb the background geometry. Even if these fluctuations are heavy
today, so that they have not been yet observed at accelerators, they could have
had a mass smaller than the expansion parameter $H$ during inflation. This
is particularly true if the the expansionary rate itself gives a
negative contribution to their masses, as we discussed above. If this
is the case, the fluctuations are amplified to a classical level
during the inflationary expansion.
These fluctuations can arise independently on
slow roll inflaton field.
They can be generated 
even in absence of an inflaton field (if, for example, inflation is
due to some metastable form of vacuum energy) since they are
``supported'' by the bulk dynamical fields responsible for the
stabilization of ${\cal M}\,$.

As we will see Fluctuations of the internal manifold
 will lead to modulated
perturbations. The modulated fluctuations were introduced
recently as an alternative mechanism of
generation of pimordial cosmological perturbations \cite{dgz,modulated}.

 To see this, let us focus on the four dimensional part
of the geometry.
Induced metric at $dS_4$ is
\begin{equation}\label{mmm}
\gamma_{\mu \nu}= A^2 \left( y \right) \left( 1 + 2 \, \Psi( x, y) \right) \, g_{\mu \nu} \ ,
\end{equation}
where $ g_{\mu \nu}$ is the usual four dimensional dS metric. 
Dimensional reduction leads to
appearance of the modilu field $\Psi$ 
in the four dimensional effective action.
Consider only the lightest KK mode,  after proper normalization
we write
\begin{equation}\label{action}
S=\int d^4 x \sqrt{-g}\left[R-\frac{1}{2}\Psi_{\mu}\Psi^{\mu}+\frac{1}{2}m_0^2\Psi^2
 -\Psi T^{\mu}_{\mu}\right] \ .
\end{equation}
Such an action~(\ref{action}) may appear if there are $1+3$ dimensional
submanifolds which are point-like on ${\cal M}\,$ (in which case
$\gamma_{\mu \nu}$ is the induced metric on the submanifolds). More
generally, we can also obtain it starting from the
geometry(\ref{line}) and integrating over ${\cal M}$ ( in which case
$A$ and $\Psi$ do not depend on $y\,$). The first situation is typical
of brane models, while the second is more common of general KK
theories. In both cases, we have in mind a more general action
than~(\ref{action}), which will be relevant for the dynamics of the
brane fields or of the KK modes. However, the choice~(\ref{action}) is
adequate enough for the following discussion.

The action~(\ref{action}) differs from a standard 
theory of inflation  due to the presenc of interaction term $\Psi T^{\mu}_{\mu}$.
Perturbations of  $\Psi$
 lead to rescaling of masses and couplings during inflation,
untill the moduli $\Psi$ is pinned down to its minimum where it has the 
the mass $m_0^2$.
The spatial fluctuations ${\Psi}$ of the geometry introduce a small spatial
dependency in this redefinition, so that any physical mass scale and couplings  will
be slightly different in different parts of the universe.

Inhomogeneities of the moduli field $\Psi$ are generated during inflation
if its mass $m_0$ is ligther than the Hubble parameter.
Now we recall the results of the previous section, that
 $m_0^2$ contains tachyonic contribution.
Therefore, an interesting possibilitu arises if the net value of
 $m_0^2$ is smaller than $H^2$. In this case during inflation 
we encounter generation of spatial fluctuations of $\Psi$.
In the late universe long after inflation, when $H$ significantly drops,
we expect $m_0^2$ be positive and large relatively to the temperature of the universe,
si that $\Psi$ is stabilized.

Small spatial variation of masses and coupingsm immediately after inflation,
 during (p)preheating after inflation,  will
result in a spatial dependency of the inflaton decay rate $\Gamma$
\begin{equation}
\frac{\Delta \Gamma}{\Gamma} \sim { \Psi} \,\,.
\end{equation}
This fluctuation in the decay of the inflaton give rise to the adiabatic
mode of the (standard)
perturbations: after the reheating epoch the evolution of the perturbations
in the various decay products follows the standard picture. The important
distinction is that in this case the inflaton assumes a secondary role in defining
the nature of the perturbations. 

After reheating ends and modilu $\Phi$ is trapped in its minimum,
spatial variatins of masses and couplings are erased.
However, long wavelength metric perturbations generated during the inflaton decays 
are preserved as the scalar cosmological perturbations \cite{dgz,modulated}.

There has been some activity in the computation of the spectrum of CMB
anisotropies in braneworld models. However, many of these studies do
not discuss the general set of metric perturbations in the bulk, but
rather focus on the evolution of the standard four dimensional perturbations
with the braneworld configuration as a background. On the contrary, the
mechanism outlined here exploits the presence of new perturbations,
related to the bulk dynamic, in a crucial way (notice that the conformal
mode~(\ref{action}) is absent on the standard $4$ dimensional case).
In addition, the one we have described is a very natural realization of the
general idea of modulated perturbations proposed in~\cite{modulated,dgz},
since we provide a definite candidate
(generally present in all the configurations~(\ref{blocks}) for the modulus field
controlling $\Gamma\,$).

There has been much focus
recently on so--called ``anomalies'' emerging in the observations of
LSS and the CMB~\cite{spergel} ranging from an intriguing lack of power on the
largest scales~\cite{cpkl}, the possibility of ``glitches'' in the CMB spectrum
and an indication of running of the spectral index of the primordial
scalar perturbations (for a recent review, see~\cite{cbi}), to detections of
mild non-Gaussianity and
indications of a finite topology for the $3$ spatial dimensions. The
accuracy of future observations is expected to increase greatly over
the next decade and the hope is that if any of the anomalies turn out
to be robust results they will help to constrain new physics modifying
the dynamics in the early universe.

\section{Discussion}~\label{sec:end}

A typical denominator in many extensions of the Standard Model is
the presece of extra-dimensions. In the simplest possibilities, the
extra space is a compact and static manifold ${\cal M}\,$. A stabilization
mechanism is typically required to explain why ${\cal M}$ should remain static,
while the $3$ noncompact spatial dimensions are undergoing cosmological
expansion. In particular, significant activity has been recently made trying to
reconcile this picture with a de Sitter (or quasi de Sitter) geometry for the
noncompact coordinates. Indeed, observations are telling us that the expansion
of the universe was accelerating at very early times, and it is accelerating also
at present.

The present note is focused on the effects of the inflationary expanison
on the stability of ${\cal M}\,$. We have argued that the dS curvature (in other
words, a nonvanishing expansion rate $H$) has typically the effect to
destabilize the internal space, and that any given stabilization mechanism
can be effective only up to a certain curvature. We have aimed to discuss
this effect in the most general way possible, enlighting the consequences of making the
only assumption that the geometry is of the $dS \times {\cal M}_4$ type, with an arbitrary
compact space of $d$ dimsensions, and possibly with the presence of a warp factor.
Such a set-up is also relevant for string-theory, in the supergravity limit.

To discuss the stability of the system, we have studied the most general  set
of perturbations of the $dS_4 \times {\cal M}_4$ geometry. It is very convenient to
classify the perturbations into irreducible representations of the $dS_4$ symmetry group.
The big advantage of doing so, is that modes belonging to different representations
are not coupled at the linear level. The perturbations can be devided into scalar, vecor, or tensor
modes with respect to the $dS_4$ isometries. After a general classification, we have
focused on the scalar modes, since they are ususally the most relevant one for inflationary
geometries, and since they encode the effects of the instability we want to discuss.
The linearized Einstein equations for the perturbations show that the physical 
mass squares $m^2$ of these modes acquire a negative contribution due to the $dS$ expansion,
\begin{equation}
m_n^2 \left( H \right) = - \frac{12 \, H^2}{1+d/2} \,\,.
\label{tacconc}
\end{equation}
This is true for arbitrary ${\cal M}\,$, $d\,$, and for any possible underlying
stabilization mechanism.

The presence of this tern is a signal for a possible gravitational instability of the system. Indeed,
if the whole $m_n^2$ turns out to be tachyonic, the $dS \times {\cal M}$ is unstable. Clearly,
verifying whether (and at which $H$) the instability takes place is very model dependent issue
which should be verified case by case (namely, for any given source $T^A_B \,$).
Indeed, other contributions to $m^2$ also dpend on $H$
(although in most cases only indirectly, due to the fact that these terms depend on other
background quantities, and that these quantities are related to $H$ through the background
Einstein equations). However, the contribution~(\ref{tacconc}) is {\it generally}
present, and it has to be taken always into account. Moreover, the study of several different examples,
as well as general arguments, lead us to conjecture that the term~(\ref{tacconc}) is the dominant one
at large $H\,$, and that no stabilization mechanism can be effective up to arbitrarly large dS curvature.

In this work, we have presented explicit and exact calculations in two specific examples. The first
of them is codimension one braneworld configurations with bulk scalar field(s). As shown in~\cite{FK},
in this case it is possible to derive an explicit upper bound $m^2 \leq - 4 \, H^2 + \mu_0 \left( A
\right)\,$ (where $A$ is the warp factor, see eq.~(\ref{mass3})) for the mass of the lightest eigenmode.
This bound allows to determine at which $H$ the system becomes unstable. The second example is when
${\cal M}$ is a $d-$sphere, and when the only source term isw a cosmological constant. This system
is well known to be unstable, and we have shown that the instability is precisely due
to the tachyonic nature of the scalar modes which we have identified in the general calculation.

There are also general arguments in support of the instability at high $H\,$.
The first is related to causality. If, as in the standard $4$ dimensional case, the Hubble length $H^{-1}$
has the meaning of a casual horizon, we should expect that any stabilization mechanism cannot
be at work when $H$ becomes much greater than the inverse size of ${\cal M}$ (since
different ``edges'' of ${\cal M}$ would then be casually disconnected). Such a behavior can be
found in~\cite{lpps}, where the cosmology of the Randall-Sundrum model~\cite{RS1} was
studied under the assumption of radion stabilization. It was shown 
in~\cite{lpps} that radion
stabilization cannot be imposed if the physical energy on the
hidden brane is greater than $\sim {\rm TeV^2} \, M_p^2 \,$.
The appearence of this intermediate scale is somewhat surprising, and indeed it was left unexplained
in~\cite{lpps}. However, it can be verified that it is precisely at this 
energy density that $H^{-1}$ becomes
smaller than the distance between the two branes.~\footnote{We thank Lorenzo Sorbo
for this observation.}

A second argument can be inferred from the $dS_4 \times S_d$ example. We noted that the instability
has two possible end-points. One is characterized by a shrinking internal space, with asymptotic 
Kasner solution (a similar behavior was already noted in~(\cite{branecode}) in the case of brane collisions).
The second one is instead exact $dS_{4+d}\,$, with all the compact and noncompact dimensions
expanding with the same asymprtotic rate. This suggests that the instability can be attributed to the tendency
of an inflationary expansion to homogenize and isotropize the {\it whole} geometry, in this case
by ``dragging'' also the compact space into a de--Sitter solution. This behavios, which is
a well known feature of inflationary expansions, is encountered also in many Bianchi models.

In the final part of the work, we have also commented on the possible role of
the instability in the generation of primordial perturbations. The scalar perturbations appear as a conformal
mode from the four dimensional point of view (for example, for an observer living on a $3$-brane, or
after integrating out the internal space ${\cal M}\,$). This mode is a clear signature of the extra dimesnions,
since it is absent in the standard $4$ dimensional case. If, due to the tachyonic contribution, this mode
light ($0 < m < H$) during the de Sitter stage, it will be amplified with a scale invariant spectrum
and with an amplitude which is disentangled from the standard result of slow-roll inflation. One can think of
a number of ways how the conformal perturbation could be responsible for the adiabatic mode in the later
FRW evolution. For instance, consider a mixture of fields conformally and nonconformally coupled to
the noncompact geometry. Only the nonconformally coupled fields will be sensitive to the conformal perturbations,
so that we expect that the final mode perturbation will be proportional to the amount of these fields in
the mixture. In the text we have instead focused on a much simpler possibility, related to the mechanism of
modulated perturbations~\cite{modulated,dgz}. The conformal factor can be usually rescaled away by a rescaling
of the energy scales in the $4$ dimensional theory (this is exactly what occurs in the Randall-Sundrum
model~\cite{RS1}). Hence, its fluctuations can be interpreted as fluctuations masses of
particles and rates of processes in the $4$ dimensional theory. In oarticular, this may be the origin of the fluctuations
of the decay rate of the inflaton, which is at the basis of the mechanism of modulated perturbations.

\section{Acknowledgments}

It is a pleasure to thank  Johannes Martin, 
Dimitry Podolsky, and Lorenzo Sorbo for very useful discussions.

\appendix

\section{Operators in the Linearized equations for the perturbations} \label{a1}

The linearized equations~(\ref{linpert})  for $\Psi$ have been given in the main text in terms
of the three operators $L_{1,2,3}\,$. These operators are
\begin{widetext}
\begin{eqnarray}
L_1 &=& \left(d + 4 \right) \,
\frac{{\hat \nabla}^a A}{A} \, {\hat \nabla}_a + 4
\, \frac{\Box^{\hat{}} A}{A} +
2 \, \left( d - 1 \right) \frac{\left( {\hat \nabla} A \right)^2}{A^2}
\,\,, \nonumber\\
L_2 &=& 2 \, \frac{{\hat \nabla}_a A}{A} \,
{\tilde \nabla}^\mu \nonumber\\
L_3 &=& - 2 \frac{{\hat \nabla}^a A}{A} \, {\hat \nabla}_b
- 2 \frac{{\hat \nabla}_b A}{A} \, {\hat \nabla}^a + 8 \frac{ 
{\hat \nabla}^a A {\hat \nabla}_b A}{A^2} - 4
\frac{{\hat \nabla}^a {\hat \nabla}_b A}{A} +
\delta^a_b \left[ 2 \left( d + 3 \right)
{\hat \nabla}^c A {\hat \nabla}_c +
4 \, \frac{\Box^{\hat{}} A}{A} + 2 \left( d - 1 \right)
\frac{\left( {\hat \nabla} A \right)^2}{A^2} \right] \,\,.
\end{eqnarray}
\end{widetext}
where quantities with tilde are computed from the background metric on
$x\,$, while quantity with hat from the background metric on $y\,$.


\begin{thebibliography}{99}


\bibitem{Mal}
J.~M.~Maldacena and C.~Nunez,
Int.\ J.\ Mod.\ Phys.\ A {\bf 16}, 822 (2001)
[arXiv:hep-th/0007018].


.

\bibitem{ds1}
S.~B.~Giddings, S.~Kachru and J.~Polchinski,
Phys.\ Rev.\ D {\bf 66}, 106006 (2002)
[arXiv:hep-th/0105097].



\bibitem{KKLT}
S.~Kachru, R.~Kallosh, A.~Linde and S.~P.~Trivedi,
Phys.\ Rev.\ D {\bf 68}, 046005 (2003);
S.~Kachru, R.~Kallosh, A.~Linde, J.~Maldacena, L.~McAllister and S.~P.~Trivedi,
JCAP {\bf 0310}, 013 (2003)
[arXiv:hep-th/0308055].



\bibitem{ds2}
A.~Lukas, B.~A.~Ovrut, K.~S.~Stelle and D.~Waldram,
Nucl.\ Phys.\ B {\bf 552}, 246 (1999).

\bibitem{ss}
A.~Salam and E.~Sezgin,
``Supergravities In Diverse Dimensions. Vol. 1, 2,'',


\bibitem{GW}
W.~D.~Goldberger and M.~B.~Wise,
Phys.\ Rev.\ Lett.\  {\bf 83}, 4922 (1999).

\bibitem{Cas} 
P.~Candelas and S.~Weinberg,
Nucl.\ Phys.\ B {\bf 237}, 397 (1984).

\bibitem{pp}
E.~Ponton and E.~Poppitz,
JHEP {\bf 0106}, 019 (2001).

\bibitem{radion} 
C.~Csaki, M.~L.~Graesser and G.~D.~Kribs,
Phys.\ Rev.\ D {\bf 63}, 065002 (2001)
[arXiv:hep-th/0008151].

\bibitem{kmp}
L.~Kofman, J.~Martin and M.~Peloso,
[arXiv:hep-ph/0401189].

\bibitem{KS}
L.~A.~Kofman and A.~A.~Starobinsky,
Phys.\ Lett.\ B {\bf 188} (1987) 399.

\bibitem{B}
J.~M.~Bardeen,
Phys.\ Rev.\ D {\bf 22}, 1882 (1980).

\bibitem{mfb}
V.~F.~Mukhanov, H.~A.~Feldman and R.~H.~Brandenberger,
Phys.\ Rept.\  {\bf 215}, 203 (1992).

\bibitem{bruc}
C.~van de Bruck, M.~Dorca, R.~H.~Brandenberger and A.~Lukas,
Phys.\ Rev.\ D {\bf 62}, 123515 (2000)
[arXiv:hep-th/0005032].

\bibitem{Higuchi}
A.~Higuchi,
Nucl.\ Phys.\ B {\bf 282}, 397 (1987).

\bibitem{cgr}
C.~Charmousis, R.~Gregory and V.~A.~Rubakov,
Phys.\ Rev.\ D {\bf 62}, 067505 (2000)
[{\tt hep-th/9912160}].
 
\bibitem{FK}
A.~V.~Frolov and L.~Kofman,
Phys.\ Rev.\ D {\bf 69}, 044021 (2004)
[arXiv:hep-th/0309002].

\bibitem{branecode} J.~Martin, G.~N.~Felder, A.~V.~Frolov, M.~Peloso
and L.~Kofman, Phys.\ Rev.\ D, to appear
[{\tt hep-th/0309001}]

\bibitem{sasaki} 
U.~Gen and M.~Sasaki,
Prog.\ Theor.\ Phys.\  {\bf 105}, 591 (2001)
[arXiv:gr-qc/0011078].

\bibitem{ks}
L.~A.~Kofman and V.~Sahni,
Phys.\ Lett.\ B {\bf 127}, 197 (1983).

\bibitem{bkl}
V.~A.~Belinsky, I.~M.~Khalatnikov and E.~M.~Lifshitz,
Adv.\ Phys.\  {\bf 19}, 525 (1970).

\bibitem{dh}
T.~Damour and M.~Henneaux,
Phys.\ Rev.\ Lett.\  {\bf 85}, 920 (2000)
[arXiv:hep-th/0003139].

\bibitem{lpps}
J.~Lesgourgues, S.~Pastor, M.~Peloso and L.~Sorbo,
Phys.\ Lett.\ B {\bf 489}, 411 (2000)
[arXiv:hep-ph/0004086].

\bibitem{RS1}
L.~Randall and R.~Sundrum,
Phys.\ Rev.\ Lett.\  {\bf 83}, 3370 (1999);








\bibitem{modulated}
L.~Kofman,
[{\tt astro-ph/0303614}].

\bibitem{dgz}
G.~Dvali, A.~Gruzinov and M.~Zaldarriaga,
Phys.\ Rev.\ D {\bf 69}, 023505 (2004).



\bibitem{spergel}
D.~N.~Spergel {\it et al.},
Astrophys.\ J.\ Suppl.\  {\bf 148}, 175 (2003)
[arXiv:astro-ph/0302209].


\bibitem{cpkl}
C.~R.~Contaldi, M.~Peloso, L.~Kofman and A.~Linde,
JCAP {\bf 0307}, 002 (2003)
[arXiv:astro-ph/0303636].


\bibitem{cbi}
A.~C.~S.~Readhead {\it et al.},
arXiv:astro-ph/0402359.




\end{thebibliography}
\end{document}